\documentstyle[twocolumn,aps,epsfig]{revtex}
\begin{document}                % INITIALIZE - DONT CHANGE %
%\draft
%\wideabs{
%\renewcommand{\thefootnote}{\fnsymbol{footnote}}
\author{A. I. Lvovsky\cite{Lvovsky}, H. Hansen, T. Aichele, O. Benson, J.
Mlynek\cite{Mlynek}  and S. Schiller\cite{Schiller}}
\address{Fachbereich Physik, Universit\"at Konstanz, D-78457 Konstanz, Germany}
\date{\today}
\title{Quantum state reconstruction of the single-photon Fock state}

\maketitle

\begin{abstract}

We have reconstructed the quantum state of optical pulses containing
single photons using the method of phase-randomized pulsed optical
homodyne tomography. The single-photon Fock state $|1\rangle$ was
prepared using conditional measurements on photon pairs born in the
process of parametric down-conversion. A probability distribution of
the phase-averaged electric field amplitudes with a strongly
non-Gaussian shape is obtained with the total detection efficiency of
($55\pm1$)\%. The angle-averaged Wigner function reconstructed from
this distribution shows a strong dip reaching classically impossible
negative values around the origin of the phase space.
\end{abstract}
 %}

\paragraph{Introduction}

States of  quantum systems can be completely described by their {\it
Wigner functions} (WF), the analogues of the classical phase-space
probability distributions. Generation of various quantum states and
measurements of their WFs is a central goal of many experiments in
quantum optics [1--3]. Of particular interest are quantum states
whose Wigner function takes on $negative$ values in parts of the
phase space. This classically impossible phenomenon is a signature of
highly non-classical character of a quantum state.

Quantum states containing a definite number of energy quanta (Fock
states $|n\rangle$) are paradigmatic in this respect. Their WFs
exhibit strong negativities and their marginal distributions are
strongly non-Gaussian (Fig.\,1). This property reflects the
fundamentally non-classical nature of these states as carriers of the
particle aspect of light.

\begin{figure}[h]
  \begin{center}
 \epsfxsize=2.7 in%5.0 in
 \epsfbox{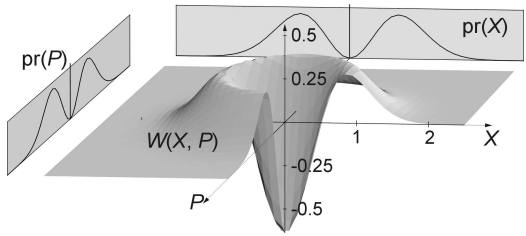}
  \end{center}

\caption{Theoretical phase space quasiprobability density (Wigner
function) of the single-photon  state $|1\rangle$: $W(X,P)={2 \over
\pi} \left( {4 (X^2+P^2)-1} \right)e^{-2 (X^2+P^2)}$. $\hat
X=(\hat{a}+\hat{a}^\dagger)/\sqrt{2}$ and $\hat
P=(\hat{a}-\hat{a}^\dagger)/\sqrt{2}i$ are normalized non-commuting
electric field quadrature observables. Single-quadrature probability
densities ({\it marginal distributions}) are also displayed.}

\end{figure}

Generation and complete measurements of the Fock states' WFs were
performed on vibrational states of a trapped ${\rm Be}^+$ ion
\cite{Be}. In the electromagnetic domain, Nogues et al.
\cite{Haroche00} recently reported the measurement of the WF of a
single-photon state in a superconducting microwave cavity at a single
point (origin) of the phase space. A full characterization of a Fock
state of the electromagnetic field has not been achieved so far.

In this paper we present a measurement of the complete
(phase-averaged) Wigner function of the propagating single-photon
state $|1\rangle$ in the optical domain. We perform a direct
measurement of the dynamical variables of the electromagnetic field,
the electric field quadratures, whereby their probability
distributions are obtained. The Wigner function is then reconstructed
from the measured distributions. This method, {\it homodyne
tomography}, has been established as a reliable technique of
reconstructing quantum states in the optical domain. Previously, it
has been applied to classical and weakly non-classical states of the
light field, such as vacuum, coherent, thermal and squeezed states,
in the continuous-wave as well as in the pulsed regime
\cite{HomoTomo}.

The main challenge associated with a tomographic characterization of
the single-photon state is the preparation of this state in a
well-defined spatio-temporal mode.
%At present, there exists no
%``photon pistol" which would emit single photons ``on command" in a
%certain spatio-temporal mode as required for homodyne quantum
%tomography. The difficulty in building such a source is that a
%microscopic event of emitting a light quantum has to be controlled
%via macroscopic means.
We solve this task by employing conditional state preparation on a
photon pair born in the process of {\it parametric down-conversion}
\cite{Mandel86,Gran86}. The two generated photons are separated into
two emission channels according to their propagation direction
(Fig.\, 2). A single-photon counter is placed into one of the
emission channels (labeled trigger) to detect photon pair creation
events and to trigger the readout of a homodyne detector placed in
the other (signal) channel \cite{Yurke87}.

\begin{figure}[h]
  \begin{center}
 \epsfxsize=2.7 in%5.0 in
 \epsfbox{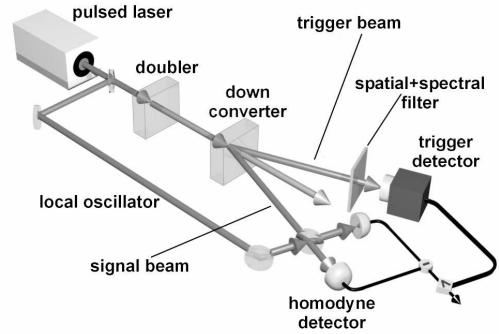}
  \end{center}

\caption{Simplified scheme of the experimental setup }\label{data}
\end{figure}

\paragraph{Theory} The process of pulsed 2-photon down-conversion produces strongly
correlated photon pairs. The generated biphoton state can be written
as

\begin{equation}\label{Psiout}
 \left| \Psi  \right\rangle=N\,\left( {\,\left| {0,0}
\right\rangle +\int {d\vec k_s\,d\vec k_t\ \Phi (\vec k_s,\vec
k_t)\,\,\left| {1_{\vec k_s},1_{\vec k_t}} \right\rangle }} \right),
\end{equation}
where $N$ is a normalization constant, $\vec k_s$  and  $\vec k_t$
denote the signal and trigger beam wave vectors, respectively, and
the function $\Phi (\vec k_s,\vec k_t)$ carries the information about
the amplitude as well as the transverse and longitudinal structure of
the photon pair generated \cite{Teich}.

A single-photon Fock state is prepared from $\left| {\Psi}
\right\rangle$ by projecting this state onto a photon count event in
the trigger beam path:
\begin{equation}\label{rho_signal}
\hat\rho_s={\rm Tr}_t\,| {\Psi}\rangle\langle {\Psi }|\ \hat\rho_t,
\end{equation}
where  $\hat\rho_t$ denotes the state ensemble selected by the
trigger and the trace is taken over the trigger states.

The trigger state ensemble $\hat\rho_t$ is determined by the spatial
and spectral filtering in the trigger channel:
\begin{equation}\label{rho_idler}
\hat\rho_t\,=\,\int {d\vec k_t\,T(\vec k_t)\,\,| {1_{\vec
k_t}\rangle\langle 1_{\vec k_t}}| },
\end{equation}
where $T(\vec k_t)$ is the spatiotemporal transmission function of
the filter. Note that although $\left| {\Psi} \right\rangle$
represents a pure state, $\hat\rho_t$ and hence $\hat\rho_s$ are
statistical mixtures. However, if sufficiently tight filtering is
applied to the trigger channel (so that $T(\vec k_t)$ is much
narrower than the spatial and spectral width of the pump beam),
$\hat\rho_s$ will approach a pure single-photon state \cite{Ou97}. In
this case, the signal photons are prepared in a relatively
well-defined optical mode suitable for homodyne detection.

It is important to understand that the ``signal beam" as shown in
Fig.\,2 is not an optical beam in the traditional sense. The
down-converted photons are in fact emitted randomly over a wide solid
angle. The optical mode of the signal state is created {\it
non-locally} only when a photon of a pair hits the trigger detector
and is registered. The coherence properties of this mode are
determined by the optical mode of the pump and the spatial and
spectral filtering in the trigger channel.

Once the approximation $\hat\rho_s$ of the Fock state is prepared, it
is subjected to {\it balanced homodyne detection}. The signal wave is
overlapped on a beamsplitter with a relatively strong local
oscillator (LO) wave in the matching optical mode. The two fields
emerging from the beamsplitter are incident on two high-efficiency
photodiodes whose output photocurrents are subtracted. The
photocurrent difference is proportional to the value of the electric
field operator $\hat E(\theta)$ in the signal mode, $\theta$ being
the relative optical phase of the signal and LO.

%To obtain a full reconstruction of a quantum state one needs to
%determine quantum noise distributions of $\hat E(\theta)\propto \hat
%X_\theta=\hat X\cos\theta+\hat P\sin\theta$ at all phase angles
%$\theta$.
%Here
%$\hat X=(\hat{a}+\hat{a}^\dagger)/\sqrt{2}$ and $\hat
%P=(\hat{a}-\hat{a}^\dagger)/\sqrt{2}i$ are normalized non-commuting
%electric field quadrature observables with $\hat{a}$ and
%$\hat{a}^\dagger$ being the annihilation and creation operators.
%Since the Hamiltonian of an optical mode is $\hat H=\hat X^2+\hat
%P^2$, $\hat X$ and $\hat P$ are analogous to the position and
%momentum operators for a particle in a harmonic potential.

For each phase $\theta$ one measures a large number $N$ of samples of
$\hat E(\theta)\propto \hat X_\theta\equiv\hat X\cos\theta+\hat
P\sin\theta$, so that their histogram (i.\,e.\,the marginal
distribution) ${\rm pr}(X_\theta)$ can be determined. The latter is
related to the WF as follows:
\begin{eqnarray}\label{prX} {\rm
pr}(X_\theta)&=&\langle X_\theta|\hat\rho_{\rm meas}
|X_\theta\rangle\\&=&\int_{-\infty}^{\infty}
W(X\cos\theta-P\sin\theta,X\sin\theta+P\cos\theta)\,dP,\nonumber
\end{eqnarray}
where $\hat\rho_{\rm meas}$ is the density matrix of the state being
measured.  The marginal distribution ${\rm pr}(X_\theta)$ can be
envisioned as a density projection of the WF $W(X,P)$ onto a vertical
plane oriented at an angle $\theta$ with respect to the plane $P=0$
(Fig.\,1). From the set of marginal distributions ${\rm
pr}(X_\theta)$ for a large number of phase angles $\theta$ the WF of
$\hat\rho_{\rm meas}$ can be reconstructed via a procedure similar to
the one used in medical computer tomography \cite{Leonhardt}.

%Quantum quadrature sampling requires direct, phase-sensitive
%measurement of the electric field which is oscillating at hundreds of
%THz. Since the bandwidth of existing detection electronics is several
%orders of magnitude lower, a special measurement method must be used,
%known as In order to enable
%homodyne detection, the optical mode of the signal photon must then
%be spatially and temporally matched to that of the local oscillator.

In a perfect experiment, $\hat\rho_{\rm meas}=|1\rangle\langle 1|$,
where the single-photon state is in the optical mode which matches
that of the local oscillator. In reality, various imperfections (such
as optical losses in the signal arm, inefficient photodiodes, dark
counts, non-ideal matching between the signal and the LO optical
modes \cite{Grangier01}) cause an admixture of the vacuum $|0\rangle$
to the measured state, so that
\begin{equation}
\hat\rho_{\rm meas}=\eta|1\rangle\langle 1|+(1-\eta)|0\rangle\langle
0|,\label{rho_meas}
\end{equation}
$\eta$ being the measurement efficiency. It is remarkable that all
these effects act upon $\hat\rho_{\rm meas}$ in a similar way, so
that their effect can be expressed in a single number $\eta$ which is
a product of efficiencies associated with individual parts of the
setup. The value of $\eta$ is crucial for this experiment as it
strongly influences the shape of the measured marginal distributions
and the reconstructed Wigner function (Fig.\,3) \cite{Yurke87}.
\begin{figure}[h]
  \begin{center}
 \epsfxsize=2.7 in%5.0 in
 \epsfbox{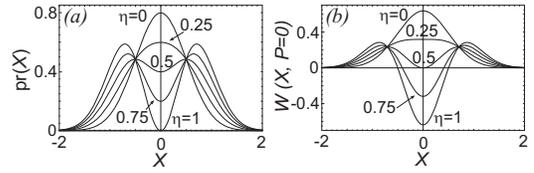}
  \end{center}

\caption{Effect of the non-perfect measurement efficiency $\eta$ on
the marginal distribution (a) and the reconstructed WF (b). For the
WF, cross-sections by the plane $P=0$ are shown. Negative values
require $\eta>0.5$.}\label{Wigner}
\end{figure}

In our experiment we used a simplified scheme in which the phase
$\theta$ varied randomly, so that we only measured a single
phase-randomized marginal distribution ${\rm pr_{av}}(X)=\langle{\rm
pr}(X_\theta)\rangle_\theta$. This does not change the measurement
result for quantum states with rotationally symmetric Wigner
functions such as those described by Eq.\,(\ref{rho_meas}). The
phase-averaged WF $W(R)$ is obtained from ${\rm pr_{av}}(X)$ via the
Abel transformation \cite{Leonhardt,Leonhardt94}:
\begin{equation}\label{Abel}
W(R)=-\frac{1}{\pi}\int_R^\infty \frac{d\,{\rm pr_{av}}(X)}{dX}
(X^2-R^2)^{-1/2}dX.
\end{equation}

From the phase-randomized marginal distributions one can also
directly infer diagonal elements $\rho_{nn}$ of the state density
matrix in the Fock basis,
\begin{equation}\label{sampling}
\rho_{nn}=\pi\int_{-\infty}^{\infty}{\rm pr_{av}}(X)\, f_{nn}(X)\,dX,
\end{equation}
where $f_{nn}(X)$ are the amplitude pattern functions \cite{patterns}
which are independent of the optical state being sampled. The
statistical uncertainty of the reconstructed $\rho_{nn}$ is
\begin{equation}\label{error}
\langle\sigma_{nn}^2\rangle={1\over N}\int_{-\infty}^{\infty}{\rm
pr_{av}}(X)\,(\pi f_{nn}(X))^2\,dX,
\end{equation}
where $N$ is the total number of field samples acquired.

\paragraph{Experimental Setup} We employed a mode-locked Ti:sapphire
laser (Spectra Physics Tsunami) in combination with a pulse picker to
obtain transform-limited pulses at 790 nm with a repetition rate 816
kHz and a pulse width of 1.6 ps. Most of the radiation was frequency
doubled in a single pass through a 3-mm LBO crystal yielding 100
$\mu$W at 395 nm and then passed on to a 3-mm BBO crystal for
down-conversion.

Down-conversion occurred in a type-I frequency-degenerate, but
spatially non-degenerate configuration, with 790-nm photon pairs
emitted at angles $\pm6.8^\circ$ with respect to the pump beam. The
BBO crystal was cut at $\theta=35.7^\circ,\ \phi=0^\circ$, so that
the direction of the walk-off of the 395 nm pump beam inside the
crystal coincided with the direction of the signal beam so as to
minimize distortions of the signal spatial mode (``hot spot"
down-conversion, \cite{HotSpot}). The short crystal lengths allowed
us to avoid group-velocity mismatch effects which would have
complicated temporal mode matching to the LO pulse.

The trigger photons passed through a spatial filter and a 0.3-nm
interference filter centered at the laser wavelength. They were then
detected by an EG\&G SPCM-AQ-131 single-photon detector (quantum
efficiency 60\%, dark count rate $<15\ {\rm s}^{-1}$) at a rate of
about $0.25\ {\rm s}^{-1}$. Such a low pair production rate
%(in comparison to the pulse repetition rate of the laser)
made the effect of Fock states with $n>1$ negligible. Precise (within
0.6 ns) gating of the count events with the laser pulses allowed us
to eliminate most of the dark counts, thereby reducing their
contribution to about 2\% of all trigger events.

We used a small fraction of the original optical pulses from the
pulse picker --- split off before the frequency-doubler --- as the
local oscillator for the homodyne system. Achieving a good spatial
and temporal {\it mode matching} between the LO and the photons in
the signal channel constituted a major challenge in this experiment
due to extremely low intensity of the field in the signal mode. To
this end, a fraction of the laser output power was directed into the
BBO crystal from the back along the trigger beam path so that it
passed through the spatial filter in the trigger channel. Inside the
crystal these alignment pulses were temporally and spatially
overlapped with the pump to produce a difference frequency (DFG)
emission into an optical mode which modeled, to a good precision,
that of the conditionally prepared signal photons. This mode was then
matched to that of the local oscillator by observing an interference
pattern between the two beams. A visibility on the level of
$v=83\pm1$\% was reached.

The method of conditional state preparation also established special
requirements for the {\it homodyne detector} electronics. The
detector needed to resolve quantum shot noises of individual laser
pulses at a 0.8-MHz repetition rate. Details on design and
performance of the homodyne system developed will be published
elsewhere \cite{HD}.

\begin{figure}[h]
  \begin{center}
 \epsfxsize=2.7 in%5.0 in
 \epsfbox{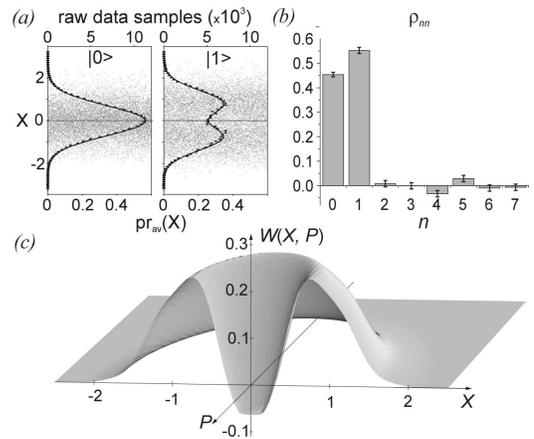}
  \end{center}

\caption{Experimental results: a) raw quantum noise data for the
vacuum (left) and Fock (right) states along with their histograms
corresponding to the phase-randomized marginal distributions; b)
diagonal elements of the density matrix of the state measured.  c)
reconstructed WF which is negative near the origin point. The
measurement efficiency is 55\%. }\label{results}
\end{figure}

\paragraph{Results and discussion} In a 14-hour experimental run about 200,000 vacuum state
and 12,000 Fock state samples were acquired (Fig.\,4a). Both data
sets were then binned up to obtain their statistical distributions. A
Gaussian distribution was fit to the vacuum state noise spectrum by
varying its X-scale and point of origin.

The best fit parameters of the vacuum state were used to scale the
Fock state data. The latter was then fit by the theoretical marginal
distribution of the ensemble (\ref{rho_meas}) to find the measurement
efficiency $\eta$. The best fit efficiency value was $\eta=0.55$.
Using the Abel transform (\ref{Abel}) the phase-randomized WF of the
observed quantum state was reconstructed (Fig.\,4c). As expected, it
exhibits negativity around the origin point, with a minimum value
$W(0,0)=-0.062$.

The diagonal elements of the density matrix have been evaluated along
with their statistical errors by applying the quantum state sampling
method as defined by Eqs.\,(\ref{sampling}, \ref{error}) directly to
the rescaled raw data. We found $\rho_{11}=0.553\pm0.013$ in
agreement with the value of $\eta$ obtained by fitting the marginal
distribution (Fig.\,4b). Substituting this quantity into
Eq.\,(\ref{rho_meas}) and calculating the corresponding value of the
WF we find $W(0,0)=-0.067\pm 0.016$, in agreement with the above
value determined from the marginal distribution. The uncertainty of
$\pm 0.016$ gives an estimation for the accuracy to which the
negative value of the Wigner function has been determined in this
experiment.

The values of $\rho_{00}$ and $\rho_{11}$ for the vacuum state were
found to be $0.9975\pm0.0029$ and $0.0021\pm0.0032$, respectively.
These quantities being equal to their ideal values within the
statistical errors indicates that no unknown conditions (e.g.
mechanical vibrations) are present that might bring the apparent
measurement efficiency above its actual value. The $systematic$
experimental errors (e.g. technical noise in the homodyne detector)
were insignificant and would result in $decrease$ of the measured
efficiency. It is thus impossible that neglecting these errors could
bring about a too optimistic estimate of $\eta$.

What are the main factors reducing the measurement efficiency? The
non-perfect match between the optical modes of LO and DFG fields
diminishes the efficiency by a factor of $v^2=0.69\pm0.02$. The fact
that the DFG wave does not perfectly mimic the conditionally prepared
mode of the single photon causes an additional reduction by
$\times0.95$. A factor of 0.90 arises from losses in the signal beam
path and non-perfect quantum efficiency of the photodiodes in the
homodyne detector. Finally, a 2\% reduction occurs due to false
trigger count events. Combining all the above factors we obtain the
upper limit estimate of the quantum efficiency as $(57\pm2)$\%, which
is in good agreement with the experimental value of $55.3$\%.

\paragraph{Conclusion and outlook} We have reconstructed the phase-averaged
Wigner function and the density matrix diagonal elements of
an optical single-photon Fock state $|1\rangle$ with a total
measurement efficiency of $55.3\pm1.3$\% using the method of
phase-randomized pulsed optical homodyne tomography. The
reconstructed WF is of non-Gaussian shape and exhibits negative
values around the origin of phase space, reflecting the strongly
non-classical character of the state $|1\rangle$ as a particle state
of the light field. Single-photon Fock states were prepared in a
well-defined electromagnetic mode by conditional measurements on
photon pairs created in the process of parametric fluorescence. The
measurement technique and error analysis were checked by performing a
simultaneous measurement on the vacuum state. Major experimental
inefficiency factors have been identified and quantified. This
experiment represents the first quantum tomography measurement of a
highly nonclassical state of the electromagnetic field.

%For the first time, classically impossible negative values of the
%phase-space quasiprobability distribution are measured without
%applying any assumptions based on quantum mechanics. This is a unique
%feature of our work in comparison to previous measurements of Fock
%state WFs \cite{Be,Haroche00}, which use quantum-mechanical analysis
%to calculate the WF of a quantum state from its experimentally
%determined density matrix elements.
%In these papers,
%the negativity of the Wigner function of the state $|1\rangle$ is not
%established experimentally, but is accepted {\it a priori} as a part
%of the theoretical background.

%

Relatively straightforward modifications of the setup would allow
tomography measurements of displaced Fock states \cite{DispFock} and
1-photon added coherent states \cite{PhotAdd}. Of special interest is
the entangled state $|0,1\rangle-|1,0\rangle$ generated when a single
photon is incident on a beamsplitter. This state can also be used to
demonstrate nonlocality of a single photon \cite{nonlocality}. In a
more distant future, arbitrary quantum states of the light field
might be generated via repeated down-conversion \cite{HH}.

We thank S. Eggert, C. Hettich and P. Lodahl for their help in
building up the experimental setup. A. L. is supported by the
Alexander von Humboldt Foundation. This work was funded by the
Deutsche Forschungsgemeinschaft.


\begin{references}
\bibitem[*]{Lvovsky} email: Alex.Lvovsky@uni-konstanz.de
\bibitem[\dagger]{Mlynek} Present address:
President, Humboldt-Universit\"at zu Berlin, D-10099 Berlin, Germany
\bibitem[\ddagger]{Schiller} Present address: Institut f\"ur
Experimentalphysik, Heinrich-Heine-Universit\"at D\"usseldorf,
D-40225 D\"usseldorf, Germany
%\bibitem{Wigner} E.P. Wigner, Phys. Rev. {\bf 40},
%749 (1932)
\bibitem{Be}  D. Leibfried {\it et al.} Phys. Rev. Lett. {\bf
77}, 4281 (1996)
\bibitem{Haroche00} G. Nogues {\it et al.} Phys. Rev. {\bf A
62}, 054101 (2000).
\bibitem{HomoTomo} D. T. Smithey {\it et al.} Phys. Rev. Lett.
{\bf 70}, 1244 (1993); G. Breitenbach, S. Schiller, and J. Mlynek,
Nature {\bf 387}, 471 (1997); M. Vasilyev {\it et al.}, Phys. Rev.
Lett. {\bf 84}, 2354 (2000)
\bibitem{Mandel86} C. K. Hong and L. Mandel, Phys. Rev. Lett. {\bf 56}, 58
(1986)
\bibitem{Gran86} P. Grangier, G. Roger and A. Aspect,
Europhys. Lett. {\bf 1}, 173 (1986)
\bibitem{Yurke87} B. Yurke and D. Stoler, Phys.
Rev. {\bf A 36}, 1955 (1987)
%\bibitem{Asp89} Aspect, A., Grangier, P. \& Roger, G. Wave particle duality for
%a single photon. {\it J. of Opt.} {\bf 20}, 119 (1989)
\bibitem{Ou97} Z. Y. Ou, Qu. Semiclass. Opt. {\bf 9}, 599 (1997)
\bibitem{Teich} A. Joobeur {\it et al.}, Phys. Rev. {\bf A 53}, 4360 (1996)
\bibitem{Leonhardt} for details on quantum tomography and inverse Radon transformation,
see U. Leonhardt, {\it Measuring the quantum state of light},
Cambridge University Press, 1997
\bibitem{Grangier01}F. Grosshans and P. Grangier, Eur. Phys. J. {\bf
D 14}, 119 (2001)
\bibitem{Leonhardt94} U. Leonhardt and I. Jex, Phys. Rev. {\bf A 49}, 1555
(1994)
\bibitem{HotSpot} K. Koch {\it et al.}, IEEE J. Quant. El. {\bf 31}, 769 (1995)
\bibitem{patterns}
G. M. D'Ariano, U. Leonhardt, and H. Paul, Phys. Rev. {\bf A 52},
R1801 (1995); U. Leonhardt and M. G. Raymer, Phys. Rev. Lett. {\bf
76}, 1985 (1996)
%\bibitem{vibrations} As observed at earlier stages of the
%experiment, such phenomena may indeed occur, e.g., due to mechanical
%vibrations.
\bibitem{HD} H. Hansen {\it et al.} {\it quant-ph}/0104084
\bibitem{DispFock} K. E. Cahill and R. J. Glauber, Phys. Rev. {\bf 177}, 1857 (1969)
\bibitem{PhotAdd} G. S. Agarwal and K. Tara, Phys. Rev. {\bf A 43}, 492 (1990);  C. T. Lee, Phys. Rev. {\bf A 52}, 3374
(1995)
\bibitem{nonlocality} S. M. Tan, D. F. Walls, and M. J. Collett, Phys. Rev. Lett. {\bf 66}, 252
(1991); K. Banaszek and K. Wodkiewicz, $ibid.$ {\bf 82}, 2009 (1999);
K. Jacobs and P. L. Knight, Phys. Rev. {\bf A 54}, 3738 (1996)
\bibitem{HH}  J. Clausen {\it et al.}, {\it quant-ph}/0007050
\end{references}
\end{document}